# Graphene Antennas: Can Integration and Reconfigurability Compensate for the Loss?

J. Perruisseau-Carrier, M. Tamagnone, J. S. Gomez-Diaz, and E. Carrasco
Adaptive MicroNano Wave Systems, LEMA/Nanolab
Ecole Polytechnique Fédérale de Lausanne (EPFL), Switzerland
julien.perruisseau-carrier@epfl.ch

*Abstract—* **In this opening presentation we will first recall the main characteristics of graphene conductivity and electromagnetic wave propagation on graphene-based structures. Based on these observations and different graphene antenna simulations from microwave to terahertz, we will discuss the issue of antenna efficiency, integration and reconfigurability, as function of the operation frequency range. While the use of graphene for antennas at microwave appears extremely limited, the plasmonic nature of graphene conductivity at terahertz frequency allows unprecedented antenna properties and in particular efficient dynamic reconfiguration.**

## I. Graphene conductivity

The use of graphene for antennas could potentially lead to very interesting features such as miniaturization, integration with graphene RF active electronics [1], dynamic tuning, and even optical transparency and mechanical flexibility.

For antenna applications, the property of interest of this truly 2D material is naturally its surface conductivity tensor [2, 3]. This conductivity tensor depends on graphene unique band structure and on a number of parameters including temperature, scattering rate, Fermi energy, electron velocity, pre-doping (not all these parameter are independent). Importantly, it can also be controlled via static electrical and magnetic field bias, thereby offering unprecedented possibilities for dynamic reconfiguration.

In practice other parameters such as defects in the potential polycrystalline nature of the graphene can further affect its properties. In general, the conductivity of graphene is very frequency-dependent, and can have completed different behavior e.g. at microwave and THz. Fig. 1 provides an example of the scalar graphene conductivity as a function of bias electric field, which essentially follows a Drude-like behavior. In the context of this paper, the following main feature of graphene conductivity should be highlighted:

- If a magnetic field bias is present or if spatial dispersion is considered, graphene conductivity is non-scalar. Therefore magnetic bias can be used to dynamically control e.g. Faraday rotation [4, 5]. However this case is not considered here.
- An applied electric field bias injects more electron or holes carriers in graphene and thereby allows the dynamic control of both real and imaginary part of the conductivity. However, as can be observed in Fig. 1, at micro or mm-wave frequencies this means controlling only the real part of the conductivity, which has obvious consequence on antenna reconfiguration possibilities as discussed below.
- Though graphene is the best electrical conductor known, it is mono-atomic and thus the surface resistance is very high compared to metals at micro and mm-waves frequency, even with the possibility of doping and electric field biasing. In this frequency range graphene is thus mostly a moderate to bad conductive surface.
- Graphene plasma roll-off frequency is lower than for nobles metals, namely in the THz range. Therefore its conductivity has a totally different behavior at THz (equivalent to a negative real permittivity for a bulk material). Plasmonic propagation can thus be supported by graphene at THz, leading to extremely interesting properties for practical applications.

Finally, note that the conductivity of graphene has been directly or indirectly measured using different methods at DC up to THz (see e.g. [4, 6, 7]).

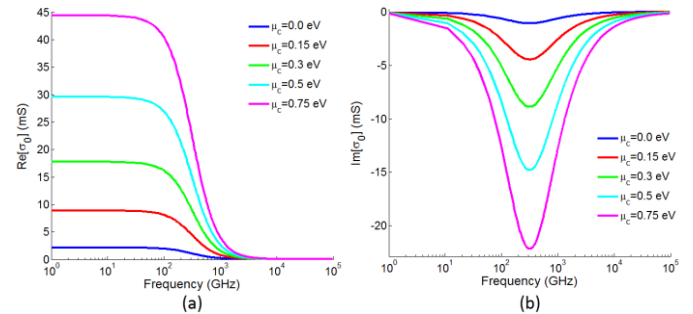

Fig. 1. Real (a) and imaginary (b) parts of graphene conductivity versus chemical potential. Graphene parameters are $\tau=0.5$ps and $T=300°$K.

## II. Graphene antennas at microwaves

Initial works on graphene microwave antennas focused on the use of graphene as a component parasitic to radiation, i.e. graphene is not supporting the main radiating currents [8, 9]. In this section, we discuss the use of graphene as an actual radiator, i.e. a platform to couple guided electromagnetic waves to free space at microwaves. Specifically, we consider a few antenna examples with different resonant frequencies and feeding mechanisms, analyzing their radiation characteristics versus graphene features.

The first study of a graphene-based patch antenna at microwaves was presented in [10]. The inset of Fig. 2a shows the antenna, designed to resonate at 2.3GHz and fed by a matched microstrip line. Importantly, we consider here a patch of highly doped graphene of excellent quality ($\tau$=0.5ps and $\mu_c$=0.25 eV). Fig. 2a and Fig. 2b show the antenna return loss and radiation efficiency, respectively. Results clearly indicate that, thought the patch is well-matched to the feeding line, the antenna radiation efficiency is low due to the intrinsic dissipation losses of graphene.

A second important observation is that applying an electric field bias to the antenna (i.e. affecting graphene chemical potential) only very slightly modifies the resonant frequency of the patch. This behavior is perfectly explained by Fig. 1, where it is observed that at microwave frequencies only the real part of the conductivity is significant and affected by the bias field. Therefore bias here allows controlling the antenna efficiency but has virtually no use for reconfiguration.

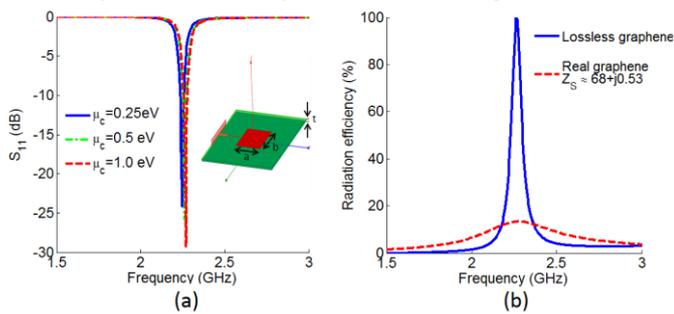

Fig. 2. Graphene-based patch antenna operating at microwaves [10]. (a) Returns loss of the antenna versus graphene chemical potential. The inset shows an illustration of the antenna. The parameters are a=b=41.08mm, thickness t=1.52 mm and dielectric permittivity $\varepsilon_r$=3.2. (b) Antenna radiation efficiency considering a patch composed of ideal lossless graphene (solid line) or of real lossy graphene (dashed line). Graphene parameters are $\tau$=0.5ps, $\mu_c$=0.25 eV and T=300°K

We present here another similar brief study related to an aperture-coupled antenna to be used in the band from 14 GHz to 16 GHz. The radiating structure is shown in Fig. 3a. In this case the ground planes and feeding microstrip line have been designed using copper, while the patch has been designed considering a surface resistance of 100Ω (corresponding to graphene parameters of $\tau$=0.5ps and $\mu c$=0.175 eV. Note that the surface reactance can be neglected at such frequencies, see Fig. 1), which can be considered as a good conductivity requiring high-quality graphene, doping, or multilayer configurations. The geometry has been optimized for obtaining at least 20dB of return loss in the whole band (50Ω reference), as can be seen in Fig. 3b. The radiation efficiency is around 8% (see Fig 3c).

With the aim of comparing the behavior of the graphene patch with the case of a good conductor patch, a simulation has been performed without modifying any of the geometric dimensions (this is of course a bad choice for the metal implementation, but allows us to draw interesting conclusions). As expected, the matching is not good, i.e. around -3.5dB at 15GHz, but the radiation efficiency is considerably better than that achieved with graphene. Observing now both effects simultaneously (mismatch and radiation efficiency), the radiated power is also better for gold than in the case of graphene despite the very suboptimal dimensions of the metal implementation (see Fig. 3d). Even more interesting, we observe that in fact removing the graphene patch allows even better radiation, i.e. that of coupling slot alone. In other words, the graphene patch with conductivity in the order of a few tens of ohms is in this case only absorbing energy and is in fact not contributing positively to the radiation mechanism.

Finally a similar study was made using a highly efficient IPD patch at 60 GHz [11]. This example is not shown here for space consideration but lead to similar conclusions.

In conclusion, for a majority of microwave antenna designs graphene is not useful: it does not provide efficient reconfiguration nor miniaturization; concerning graphene mechanical flexibility or optical transparency, they become irrelevant if some part of the antenna is made of metal or if simply suppressing the graphene allows better performance, as illustrated in the examples above. It is important to point out that these examples only considered a simple monoatomic graphene layer, and that the use of more complicated topologies, such as multilayer graphene or graphene stacks [12], could improve the radiation efficiency of this antennas. However, these solutions highly increase the fabrication complexity of the resulting devices while still providing lower radiation efficiencies than standard metallic patches without other significant benefit.

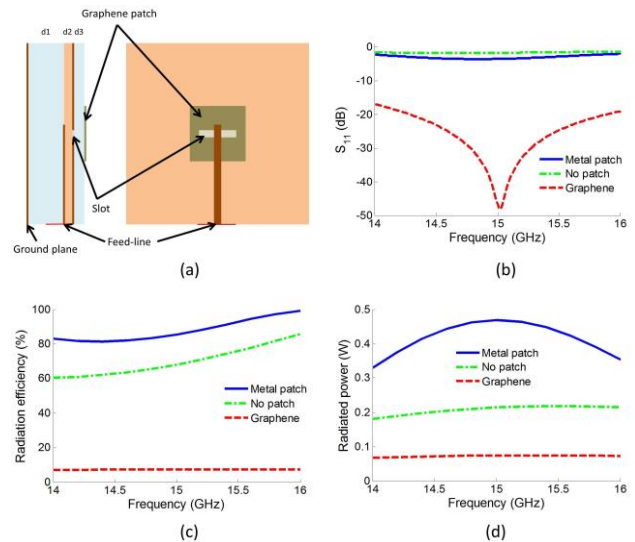

Fig. 3. Aperture-coupled antenna at 15 GHz. (a) General configuration of the antenna. The substrates employed are Rohacell©, with $\varepsilon_r$=1.05 and tan&=0.0017, for d1 and d3 (with thickness d1=5mm and d3=0.32mm) and Rogers R04450B, with $\varepsilon_r$=3.38, tan&=0.004 and thickness 0.31mm, for the dielectric d2. The dimensions of the square patch and the rectangular slot are 6.11x6.11 mm$^2$ and 4.20x0.82 mm$^2$, respectively. (b) Return loss of the antenna. (c) Radiation efficiency. (d) Radiated power for an input power of 1W.

### III. Graphene antennas at Terahertz

The inductive behavior of graphene conductivity (see Fig. 1) in the THz band enables the propagation of transverse magnetic (TM) plasmonic surface modes [13]. These modes are not available in the microwave range, where the real part of the conductivity dominates. Properly sized graphene patches act as THz plasmonic resonators [14]. However, structures such as the one presented in [14] behave merely as scatterers for THz waves, and cannot be considered as radiating antennas able to couple lumped THz sources (such as THz photomixers [15]) to free space propagation.

An actual radiator based on plasmonic modes on graphene can be obtained connecting the THz source to two graphene patches [12], as shown in Fig. 4a and Fig. 4b. Alternatively, two stacked graphene sheets separated by an insulating layer can be used instead of a single patch (Fig.4c). The latter setup allows a dynamical reconfiguration based on the electric field effect occurring when a DC voltage is applied between the stacked sheets. This antenna exhibits several very interesting features. First, the real part of the input impedance (Fig. 5) at resonance is very high, providing good matching to THz photomixers. Secondly, the resonant frequency cane be tuned in the wide range 0.5 - 2 THz and exhibits a remarkable uniform behavior for low chemical potential values. Finally, though the input impedance resembles the one of a standard resonating dipole, the antenna is highly miniaturized. This can be explained employing a TL model (Fig. 4d) based on plasmonic propagation [12], which relates the small size of the resonating structure to the low phase velocity of plasmonic modes.

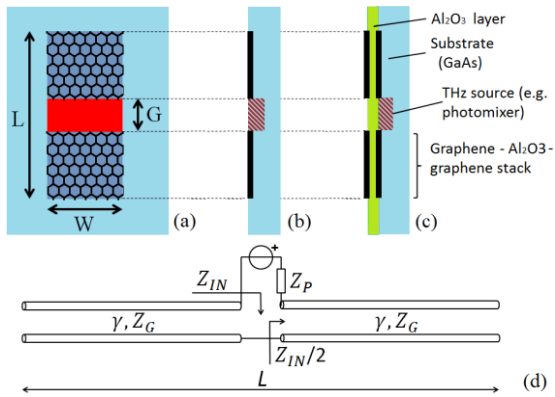

Fig. 4. THz graphene plasmonic dipole antenna. (a) Top view. (b) Cross section (single layer antenna). (c) Cross section (stacked antenna). (d) TL model.

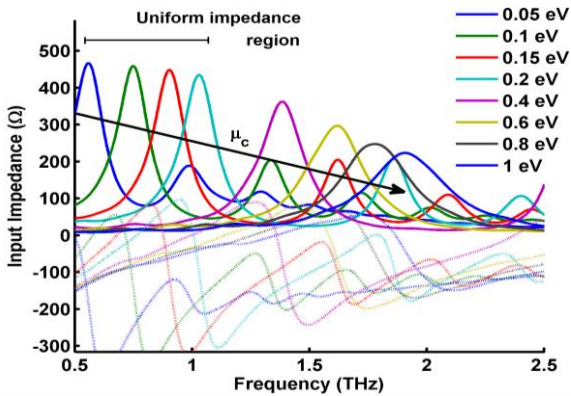

Fig. 5. Input impedance of a stacked antenna having as parameters L=22μm, W=7μm, G=2μm. The uniform impedance behavior for low chemical potential is highlighted.

The total efficiency of the antenna when operating with a source having an internal impedance of 10 kΩ exceeds 6% for chemical potentials higher than 0.6 eV. Typical metal resonant antennas can reach 20% of total efficiency, but they are larger and cannot be frequency tuned. However it is extremely unlikely that the efficiency of graphene antenna will supersede that of their metal counterparts, despite the statement in [14]. The above results are computed assuming a graphene relaxation time $\tau$ of 1 ps at room temperature, corresponding to high quality exfoliated graphene [16]. Importantly, this choice is consistent with the maximum size of graphene flakes that can be produced by exfoliation.

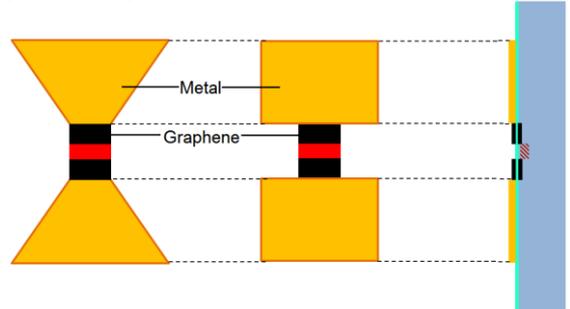

Fig. 6. Examples of metal-graphene hybrid antennas.

Finally, the efficiency of metal antennas can be combined with the input impedance tuning capabilities of graphene in hybrid graphene-metal implementations [17]. Fig. 6 shows some examples of possible topologies that can be used to achieve these improvements. In all cases, graphene is inserted between the source and the metal structure; frequency tunable Fabry-Pérot resonances of plasmonic modes on graphene provide high impedance peaks and thus good matching, while metal structures improve radiation efficiency by increasing the total size of the antenna.

The feasibility of using graphene in terahertz reflectarrays has been recently demonstrated in [18] with promising results in terms of bandwidth, low cross-polarization and grating lobe suppression, as a result of the reduced inter-element spacing provided by plasmonic propagation. Moreover, graphene conductivity can be controlled by applying an external electric field and therefore controlling the phase of the reflected field at each reflectarray element. This makes it possible to efficiently scan or reconfigure the beam [19]. Fig. 7 shows the phase of the reflection coefficient provided by a graphene square patch under normal incidence. The conductivity of the graphene patch has been computed assuming a temperature of 300K and a typical relaxation time of 1ps, easily achieved with exfoliated graphene and an optimistic value for chemical vapor deposition. The chemical potential has been varied from 0eV to 0.52eV, a very realistic value for the biasing. Note that at least 300° of phase range can be obtained in a wide band by varying the chemical potential, allowing the design of good performance reflectarray antennas. The total loss of the element in the whole bandwidth is between 0.5 dB and 6 dB, a very promising value for a controllable element at THz. Finally Fig. 8 shows the theoretical radiation patterns for a scanned-beam reflectarray with 25448 elements, at 1.3THz.

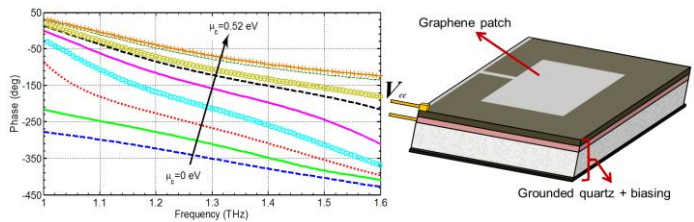

Fig. 7. Phase reconfiguration for a fixed-size square patch made of graphene, varying the chemical potential.

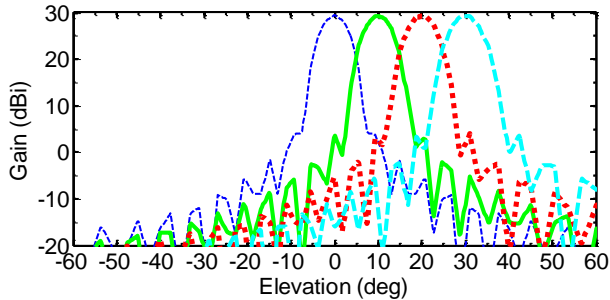

Fig. 8. Expected radiation patterns for a scanned-beam reflectarray, at 1.3THz, using graphene.

## IV. Conclusions

At microwave and mm-wave frequencies, graphene conductivity is essentially real and the electric field bias allows controlling this resistivity over a certain range. Therefore here field effect mostly allows a limited improvement in the loss but does not provide an efficient mean for reconfiguration.

Though obviously a few antenna examples as mentioned here do not exclude the possibility of better designs or more advanced fabrication (multiplayer, etc.), it is believed that graphene antennas at microwaves will generally suffer from low to very low efficiency, without practical reconfiguration capabilities. In some examples provided here we showed that even quite good graphene conductivity can lead to antennas which would perform better without graphene at all, if carefully considering return loss and radiation efficiency, which invalidates the motivation of transparency or mechanical flexibility of graphene. On the other hand graphene switches [20] might be used within a metal antenna for real reconfiguration purpose, but performance would be below that of metal MEMS for instance.

The picture is quite different at terahertz (more precisely above the plasma roll-off frequency, which essentially depend on the scattering rate), as a result of the plasmonic nature of the imaginary conductivity allowing so-called plasmonic modes. Efficient reconfiguration is now possible since electric field effect strongly affect the dominant imaginary part of the conductivity. Here it was shown that this can allow efficient control of the resonant frequency of a dipole-like antenna, or beamscanning in a reflectarray system. These features are all the more interesting that other technologies have important limitations at terahertz; graphene can provide a low-loss and simple mean for dynamic reconfiguration at such frequencies.

## References

[1] Y. M. Lin, K. A. Jenkis, A. Valdes-Garcia, J. P. Small, D. B. Farmer, and P. Avouris, "Operation of graphene transistors at gigahertz frequencies," *Nano Lett*. 9, 422–426 (2009).

[2] V. P. Gusynin, S. G. Sharapov, and J. P. Carbotte, "Magnetooptical conductivity in graphene," *Journal of Physics: Condensed Matter*, vol. 19, no. 2, p. 026222, 2007.

[3] G. W. Hanson, "Dyadic green's functions for an anisotropic non-local model of biased graphene," *IEEE Transactions on Antennas and Propagation*, vol. 56, no. 3, pp. 747–757, March 2009.

[4] I. Crassee, J. Levallois, A. L. Walter, M. Ostler, A. Bostwick, E. Rotenberg, T. Seyller, D. van der Marel, and A. B. Kuzmenko, "Giant Faraday rotation in single and multilayer graphene," *Nat. Phys*. 7, 48–51 (2010).

[5] A. Fallahi and J. Perruisseau-Carrier "Manipulation of Giant Faraday Rotation in Graphene Metasurfaces", *Applied Physics Letters*, vol. 101, n. 23, p. 231605, 2012.

[6] M. Lian, Z. Wu, L. Chen, L. Song, P. Ajayan, and H. Xin, "Terahertz characterization of single-walled carbon nanotube and grapheneon-substrate thin films," *IEEE Trans. Microwave Theory Tech*. 59, 2719–2725 (2011).

[7] J.S. Gomez-Diaz, J. Perruisseau-Carrier, P. Sharma, A. M. Ionescu, "Non-Contact Characterization of Graphene Surface Impedance at Micro and Millimeter Waves", *Journal of Applied Physics*, 111, 114908, 2012.

[8] Y. Huang, L. Wu, M. Tang, and J. Mao, "Design of a Beam Reconfigurable THz Antenna with Graphene-based Switchable High-Impedance Surface", *IEEE Transactions on Nanotechnology*, vol. 11, n. 4, pp. 836-842, July 2012.

[9] M. Dragoman, A.A. Muller, D. Dragoman, F. Coccetti, and R.Plana, "Terahertz antenna based on graphene", *Journal of Applied Physics*, 107, 104313, 2010.

[10] J. S. Gómez-Díaz and J. Perruisseau-Carrier, "Microwave to THz Properties of Graphene and Potential Antenna Applications", International Symposium on Antennas and Propagation (ISAP12), Nagoya (Japan), 2012.

[11] D. Titz, A. Bisognin, F. Ferrero, C. Laporte and H. Ezzeddine, "60 GHz Patch Antenna using IPD Technology", Antennas and Propagation Conference (LAPC), 2012 Loughborough.

[12] M. Tamagnone, J. S. Gomez-Diaz, J. R. Mosig, and J. Perruisseau-Carrier, "Reconfigurable terahertz plasmonic antenna concept using a graphene stack," *Applied Physics Letters*, vol. 101, pp. 214102-4, 2012.

[13] M. Jablan, H. Buljan, and M. Soljačić, "Plasmonics in graphene at infrared frequencies," *Physical Review B*, vol. 80, p. 245435, 2009.

[14] I. Llatser, C. Kremers, A. Cabellos-Aparicio, J. M. Jornet, E. Alarcón, and D. N. Chigrin, "Graphene-based nano-patch antenna for terahertz radiation," *Photonics and Nanostructures - Fundamentals and Applications*, vol. 10, pp. 353-358, 2012.

[15] I. S. Gregory, C. Baker, W. R. Tribe, I. V. Bradley, M. J. Evans, E. H. Linfield, A. G. Davies, and M. Missous, "Optimization of photomixers and antennas for continuous-wave terahertz emission," *Quantum Electronics, IEEE Journal of*, vol. 41, pp. 717-728, 2005.

[16] A. S. Mayorov, *et al*. "Micrometer-Scale Ballistic Transport in Encapsulated Graphene at Room Temperature," *Nano Letters*, vol. 11, pp. 2396-2399, 2011/06/08 2011.

[17] M. Tamagnone, J. R. Mosig, J. Perruisseau-Carrier, J. S. Gomez-Diaz, "Hybrid Graphene-Metal Reconfigurable Terahertz Antenna," International Microwave Symp. (IMS2013), Seattle, 2013.

[18] E. Carrasco, J. Perruisseau-Carrier, "Reflectarray Antenna at Terahertz Using Graphene", *IEEE Antennas and Wireless Propag. Lett*., vol. 12, pp.253-256, 2013.

[19] E. Carrasco, M. Tamagnone and J. Perruisseau-Carrier, "Tunable Graphene Reflective Cells for THz Reflectarrays and Generalized Law of Reflection", Applied Physics Letters, 102, 104103 (2013)

[20] P. Sharma, J. Perruisseau-Carrier and A. M. Ionescu, "Nanoelectromechanical Microwave Switch Based on Graphene", Ultimate Integration on Silicon (ULIS 2013), Warwick, UK